\documentstyle[aps,twocolumn]{revtex}
\begin{document}
\title{Bose--Einstein solitons in highly asymmetric traps}
\author{V\'{\i}ctor M. P\'erez-Garc\'{\i}a}

\address{Departamento de Matem\'aticas, Escuela
 T\'ecnica Superior de Ingenieros Industriales\\
Universidad de Castilla-La Mancha, 13071 Ciudad Real, Spain}

\author{Humberto Michinel}

\address{Departamento de Ingenieria
Mec\'anica, Naval y Oce\'anica\\
Escuela Polit\'ecnica Superior, Universidade de A Coru\~na ,
15706 Ferrol, Spain}

\author{Henar Herrero}

\address{Departamento de Matem\'aticas, Facultad de CC. Qu\'{\i}micas \\
Universidad de Castilla-La Mancha, 13071 Ciudad Real, Spain}

\maketitle

\begin{abstract}
We obtain analytic solutions to the Gross-Pitaevskii equation
with negative scattering length in highly
asymmetric traps. We find that in these traps the
Bose--Einstein condensates behave like quasiparticles and do not expand
when the trapping in one direction is eliminated.
The results can be applicable
to the control of the
motion of Bose--Einstein condensates.
\end{abstract}


\date{\today}

\narrowtext

\section{Introduction}

The recent experimental realization of 
Bose--Einstein condensation (BEC) in ultracold atomic
gases, \cite{Science,Hulet} has triggered the theoretical 
exploration of the properties of Bose gases. Specifically
there has been a great interest in the developement of 
applications which make use of the properties of this 
new state of matter. Perhaps, the 
recent development of the so--called atom laser 
\cite{atomlaser} is the best example of the interest of 
these applications. 

 The current model used to describe 
a system with a fixed mean number $N$ of weakly interacting 
bosons, trapped in a
 parabolic potential $V(r)$ is the 
following Nonlinear Schr\"odinger Equation (NLSE) (which in 
this context is called the Gross--Pitaevskii equation 
(GPE))
 \begin{equation}
\label{pura}
i \hbar \frac{\partial \psi}{\partial t} =  
-\frac{\hbar^2}{2 m} \nabla ^{2}\psi + 
V(r)\psi + U_0 |\psi|^2 \psi,
\end{equation}
which is valid when 
the particle 
density and temperature of the condensate
are small enough. Here $U_0 = 4 \pi \hbar^2 a/m$
 characterizes the
interaction and is defined in terms 
of the ground state scattering length 
$a$. 
The normalization for 
$\psi$ is $N = \int |\psi|^2 \ d^3 \vec{r},$
and the trapping potential is given by

\begin{equation}
\label{parabolic}
V(\vec{r}) = \frac{1}{2} m\nu^2 \left( \lambda_x^2 x^2 +
\lambda_y^2 y^2 + \lambda_z^2 z^2 \right),\
\end{equation}

$\lambda_\eta, \ (\eta = x,y,z)$ being, as usual, constants describing
the anisotropies of the trap \cite{Dalfovo}.
In real experimental systems the geometry of the trap
imposes the condition $\lambda_x=
\lambda_y=1$. $\lambda_z=\nu_z/\nu$
is the quotient between the frequency
 along the $z$-direction $\nu_z$
and the radial one $\nu_r \equiv \nu$.
Eq. (\ref{pura}) is strictly valid in the $T=0$ and low density
 limit, but has
been validated by different ways for the current experimental systems,
 e.g. by the comparison of the experimental
\cite{expfreq} and theoretical low energy excitation spectra of the condensates 
\cite{theorfre,Rupecht}. Recent theoretical work extends the applicability 
of the GPE to the high density limit \cite{Gard,Ziegler}.

When $a>0$ the interaction between the particles
in the condensate is repulsive, as in most current BEC experiments \cite{Science,expfreq,laser,cigar}. 
In opposite case ($a<0$), the interaction is attractive \cite{Hulet,Hulet2}.

 Although the GPE is widely accepted as a valid model 
for the dynamics of the BEC at $T \simeq 0$ K, 
the knowledge of the dynamics of the 
condensates is scarce since the GPE is non-integrable and 
explicit solutions are not known. In the positive
scattering length case Eq. (\ref{pura})
has been solved numerically for cylindrically symmetric systems
and analytically some work has been done 
in the framework of the Thomas-Fermi 
approximation \cite{Dalfovo,Rupecht,numerics}.
The negative scattering length 
case is mostly unexplored except some numerical 
results \cite{Dalfovo}. Other approach to the dynamics of the 
condensate is the time dependent variational technique \cite{Victor2}
which assumes a fixed profile 
and computes the evolution of some parameters such as the width, etc., by
variational techniques

An important fact related to negative 
scattering length condensates is that stable solutions 
to Eq. (\ref{pura}) exist only under certain 
conditions on the number of particles and the size 
of the trap \cite{Victor2,Vazquez,mathemat,chinos}. 
When those conditions are not satisfied the condensate is 
unstable and destroyed by the collapse phenomenon 
because the density $|\psi|^2$ increases up to a point 
where nonlinear 
losses (not included in Eq. (\ref{pura})) become dominant. 
So, to have a {\em large} stable negative 
scattering length condensate collapse must be
avoided. Having larger condensates is important 
to get better experimental observations of BEC. The reason
the critical number of particles that can be put 
in the condensate without collapse is very small 
for current experimental parameters and thus, it is
difficult to perform accurate measurements and to obtain
experimental data of the condensation process. Other reason for the 
interest of large condensates is their future practical
applications (atom interferometers, atom clocks, etc.), where 
coherent atom clouds as large as possible are necessary. 

In this paper we concentrate on the analysis of
negative scattering length
condensates in cigar-shaped traps. 
We present a new class of soliton--like stable solutions 
which can be of interest in the applications 
of these coherent atom aggregates. 

\section{Derivation of the model equations}

From now on we will study the solutions 
 of Eq. (\ref{pura})
in cylindrically symmetric parabolic traps (\ref{parabolic}). 
 More explicitly we will consider cigar--like 
condensates, i.e., the case in which the trapping potential 
in $s$ is much weaker than the trapping potential 
in $\rho$; mathematically $\lambda_z \ll 1$. 

Let us make the 
change of variables: $\tau = \nu t, a_0 \rho = r, a_0 s = z$ and 
$ Q = - 8 \pi a N/a_0$ where $a_0 = \sqrt{\hbar/m\nu}$ is the size of the
ground state solution of the noninteracting GPE with a harmonic potential
of frequency $\nu$ (except for a $\sqrt{2}$ factor). Let us also 
define a new wavefunction as $u(\rho,s,
\tau) = \psi(r,z,t) \sqrt{a_0^3/N}$, then
equation ($\ref{pura}$) reads 

\begin{equation}
\label{NLS3}
i \frac{\partial u}{\partial \tau} = \left[
-\frac{1}{2} \nabla ^{2} + 
 \frac{1}{2} \left( \rho^2 + \lambda_z^2 s^2 \right) - \frac{Q}{2} |u|^2 
\right] u,
\end{equation}

with the normalization condition $\int \left| u \right|^2 d\vec{r}  = 1 $. 

The solution of this nonlinear 
partial differential equation is a challenging problem and no explicit
solutions are known. However due to the different interaction 
scales involved in our particular problem it is possible to find approximate
(but very accurate) analytic forms for the ground state solutions of Eq. (\ref{NLS3}). A detailed analysis using multiscale expansions is done in 
the Appendix. Here we will derive the ground state solution by simple physical arguments.

We will first assume that it is possible
factor the solution of Eq. (\ref{NLS3}) 
as 
\begin{equation}
\label{factor}
u(\rho,s,\tau) = \phi(\rho)
\xi(s,\tau).
\end{equation}
Then, $\phi$ satisfies the 
following equation
\begin{equation}
\label{prima}
- \frac{1}{2} \nabla_{\perp}^2 \phi + 
\frac{1}{2} \rho^2 \phi = \nu_{\rho} \phi,
\end{equation}
Eq. (\ref{prima}) is a well-known eigenvalue problem, the two dimensional 
isotropic harmonic oscillator. Its {\em ground state} solution is 
\begin{equation}
\label{Gaussian}
\phi_0(\rho) = e^{- \rho^2/2}. 
\end{equation}
Multiplying Eq. (\ref{NLS3}) 
by $\phi^*$ and integrating to eliminate the $\rho$ dependence
we find 
\begin{equation}
\label{NLS1}
i \frac{\partial \xi}{\partial \tau} = -\frac{1}{2} 
\frac{\partial^2
\xi}{\partial s^2} - \frac{Q}{4} |\xi|^2 \xi + 
\frac{1}{2} \lambda_z^2 s^2 \xi + 
\nu_{\rho} \xi,
\end{equation}
where the additional factor $1/2$ in the nonlinear term
comes from the quotient 
$\int_0^{\infty} |\phi_0|^4 \rho d\rho / \int_0^{\infty}
 |\phi_0|^2 \rho d\rho =1/2$.

Finally, let us make the change $\varphi(s,\tau) = \xi(s,\tau)
 e^{i \nu_{\rho} \tau}$
to obtain 
\begin{equation}
\label{NLS2}
2 i \frac{\partial \varphi}{\partial \tau} + \frac{\partial^2
\varphi}{\partial s^2} + \lambda_z^2 s^2 \varphi -
 \frac{Q}{2} |\varphi|^2 \varphi = 0.
\end{equation}
This equation is a 1-dimensional NLSE, that in the $\lambda_z=0$ 
case can be integrated by the inverse scattering 
technique \cite{Zaharov}. When $\lambda_z=0$, Eq. (\ref{NLS2}) 
has stationary normalized single-soliton solutions of type 
\begin{equation}
\varphi(s) = \frac{\sqrt{Q}}{4\pi} 
\text{sech}\left(\frac{Qs}{8 \pi}\right), 
\label{soliton1}
\end{equation} 
From here and using the Galilean invariance of the 1d-NLSE it is possible
to find travelling soliton solutions that propagate without 
distortion.

 The width of the cloud in the $s$ direction 
is related to the nonlinear coefficient 
through the relation 
\begin{equation}
\label{width}
W_s = \sqrt{<s^2>_{\varphi}} = 4\pi^2/\left(
Q \sqrt{3} \right).
\end{equation}
 This fact 
is remarkable and means that condensates with a small number of particles
($Q$ is proportional to $N$) 
would be very long, while condensates with more particles would be
 shorter. If the number of particles is large enough the condensate 
is unstable and collapse occurs. 

To justify approximation (\ref{factor}) let us note that in
the transverse direction the 
trapping potential and nonlinear force tend to compress the wavepacket
competing against the linear dispersion effect provided by the kinetic 
energy term. On the other hand the trapping force in the $s$-direction 
has been removed so that along that axis there is only a competition 
between the nonlinear attraction and the dispersion. When the main force
on the transverse direction is the one caused by the trapping potential
the approximation will be justified. To check it let us compare both 
potentials $H_{\text{trap}} = 
\frac{1}{2} \rho^2$ and 
 $H_{\text{self-int.}} = \frac{1}{2} Q |u|^2$ for the soliton solution.  
Their ratio is a function $q(\rho,s)$ 
given by
\begin{equation}
\label{ratio}
q(\rho,s) = \frac{16 \pi^2 \rho^2 e^{\rho^2}}{Q^2 \;
\text{sech}\left(Qs/8\pi\right)}.
\end{equation}
Since $\text{sech}(x) \leq 1/2$, evidently when 
$32 \pi^2/ Q^2 \gg 1$ then $q(\rho,s) \gg 1$ except for
very small values of $\rho$. When this condition is satisfied, the 
parabolic potential dominates over the self-interaction, and then the 
only effect of the nonlinear term on the transverse shape is to 
provide small shape corrections near the center of the trap, which is 
the place where the parabolic potential is lower and the nonlinear term  more relevant. 

 To see whether the soliton solutions really exist 
we have computed numerically the
ground state solution of (\ref{NLS3}) for different values
of $\lambda_z$. In the non interacting limit (small $Q$)
the solution is given by 
\begin{equation}
\label{gaussian}
u(\rho,s) = \lambda_z^{1/4} \pi^{-3/4} 
\exp\left(-\rho^2/2-s^2/2\right).
\end{equation}
 Decreasing $\lambda_z$ and increasing $Q$
preserving also
the condition $q(\rho,s) \gg 1$ we should obtain the 
soliton solutions (\ref{soliton1}). To compute the ground state
solutions we have used the 
steepest descent method described in \cite{Dalfovo} to minimize
the Hamiltonian 
\begin{equation}
\label{Ham}
H = \int d\vec{r} \left[ \left|\nabla u\right|^2  + \left(\rho^2 
+ \lambda_z^2 s^2 \right) \left| u \right|^2 - \frac{Q}{2} \left|
u\right|^4 \right]
\end{equation}
over a discrete lattice, where the solution is defined.
In Fig. 1 we plot sections of the fundamental state $u(\rho,s)$
for different values of the trapping potential in $s$ 
(parametrized by the value of $\lambda_z$) and $Q=5$. 
As the $\lambda_z$ value decreases 
the solutions get wider, but when  
$\lambda_z = 0$ the solutions do not widen indefinitely. 
The profile is then very close to
 the one defined by Eq. (\ref{soliton1}), while the 
transverse profile is Gaussian as Eq. (\ref{Gaussian}) predicted.
For $Q=10$ it is seen in Fig. 2 that the $\lambda_z=0$ solutions
are not so close to the profiles predicted by Eq. (\ref{soliton1}). 
This is because when $Q$ is large the approximation involved in 
the derivation of Eq. (\ref{soliton1}) is not valid and 
the nonlinear energy term is comparable to the transverse 
harmonic trapping energy. However, it is striking that even 
for this large $Q$ case the differences between the numerically
calculated profile and the solitonic profile are less than 10\% .
In this case it is seen that compact 
solutions exist when the trapping in $z$ is absent. 

On the other hand, when a strong trapping potential is applied 
in the $s$-direction (stronger than the nonlinear self-interaction term) 
the numerical solution is close to the exact Gaussian ground state solution
(\ref{gaussian}), and the effect of 
the nonlinear term is only an enhanced compression of the solution near the 
center of the trap. This phenomenon is seen in Fig. 1 (plot with
$\lambda_z=0.4$).

The existence of atomic solitons has been put forward 
in \cite{Zhang} in the context of the motion of an atom beam in the field of
a travelling wave laser and in a similar context in \cite{Lenz}. 
In those papers however the trap effect was not considered and the validity 
of the transition to 1-D equations was not studied. 

\section{Control of the condensate motion}

Thus, we have numerically established the existence of 
localized solutions  and obtained their analytical form
when the trapping in $s$ is eliminated (provided the number 
of particles $N \propto Q$ is small enough to avoid collapse). 
Also there exist 
travelling solutions of this type that could propagate without
distortion. Now it is interesting
to study the response of the center of mass of the condensate to an external 
potential to because it could allow to 
control the motion of the condensate. 

Let us then study the evolution of the center of 
mass of a condensate
governed by Eq. (\ref{pura}) 
in an {\em arbitrary} external potential $V(x,y,z)$ 
(the following results are valid for any potential not only 
for {\em parabolic traps}). Defining 
\begin{equation}
\vec{X} = \int d^3 \vec{r} \left| \psi \right|^2 \vec{r}, 
\end{equation}
and computing its time derivatives using Eq. (\ref{pura}) we find
$d \vec{X}/dt = < {\cal{P}} > $, where ${\cal{P}}$ is the usual momentum operator, 
${\cal{P}}=-i\hbar \nabla $
and 
\begin{equation}
\label{Ehr}
m \frac{d^2 \vec{X}}{dt^2} = - < \nabla V >,
\end{equation}
which is the Ehrenfest theorem of Quantum Mechanics. 
Eq. (\ref{Ehr}) means that 
this theorem is still valid for the GPE so that the center of mass
of the wavepacket behaves like a classical particle.
It is possible to check the validity of (\ref{Ehr}) for 
more general NLS equations (i.e., more general nonlinear terms), 
a fact which is not well--known \cite{Alcaine}. 

This result implies that one could manipulate a 
condensate by using an external potential  as is known 
for the 1d-NLSE \cite{potenti}. Joining this result 
with the previous one, i.e., the existence of localized solutions,
we find a way to control the motion of a negative scattering 
length condensate: just relax the trap
in one direction and apply an external force along that axis, the 
condensate will respond by preserving its shape and moving like a 
classical particle. Of course the external force should be 
smoothly varying since the localized
solutions have been derived in the limit where no forces are present
\cite{import}. 

It is not strictly true that a condensate would respond as a whole 
to the external force. It is well known \cite{Zaharov} that any initial 
data evolving following (\ref{NLS2}) decomposes into solitons and ``radiation". 
So, it should be simple to find experimentally these solitonic objects by just 
adiabatically relaxing the trap and applying an external potential. Doing so 
radiation (which in these context means some free atoms) would be generated and 
some solitons obtained. Only if the initial data is already a soliton there would 
not be a breaking of the initial data into a soliton train plus radiation. 

Concerning the motion of a soliton wavepacket in a highly asymmetric trap (with 
$\lambda_z$ small but different from zero) it must be said that there are no completely stationary
solutions as shown by \cite{Moura}. However there would be a quasi-stationary solution
with gaussian tails in the $s\rightarrow \infty$ and near-sech profile in the $s=0$ region
as recently proposed in \cite{Turitsyn}. If the number of particles is large enough so that
the soliton size is small compared with the scale of variation of the potential it will be expected a smooth motion of the ``soliton" towards the ``boundary". However, if the soliton
size and trap size are of the same order of magnitude there will be a competition of 
both scales which can result in cuasiperiodic motion or even into chaotic motion as 
discussed in \cite{Perez2}.

Another ``tool" to control the motion of the condensate could be a laser field as has been put forward in \cite{Zhang,Dyrting} but in those cases 
the interaction between the trasverse 
laser field and the atoms should be carefully considered. 

\section{Application to Lithium condensates}

Let us analyze the relevance of our results for the 
Lithium Bose--Einstein condensates \cite{Hulet,Hulet2}. Following Ref. 
\cite{Hulet2} we will take as parameters $a = -14.5$ \AA and the 
usual trapping potentials for the cigar-trap that are about $\nu = 150$ Hz
corresponding to $a_0 \simeq 3 \mu$m. 

To ensure the validity of Eq. (\ref{factor})
it is necessary that $W_s \gg 1$ and then 
we find that $ N \sim 300$. However in Fig. 2 it is seen how 
even in the case $Q = 10$ ($N \simeq 900$) the differences between the
soliton profile and the real ground state are small. 

Another interesting limit corresponds to collapse. In principle one would
expect that the cigar--like trap would allow a larger number of particles
to be put before collapse occurs. 
The physical reason is that 
keeping free the condensate in one 
spatial direction collapse would not occur along that axis 
but through compression of the orthogonal (transverse) directions, 
which are smaller and thus `feel' stronger interactions. 
This means that the system would behave in a two-dimensional like manner
and then the collapse conditions should be less severe \cite{Victor2,Vazquez}. 

 To test this hypothesis we have performed simulations of the largest $Q$ value
allowed using the same code as for the computation of the ground state. The upper
limit found for the cigar-like trap is $Q = 17$ corresponding to $N \simeq 1500$. 
This number is somewhat lower than the gaussian bound given in \cite{Victor2} 
which is $Q = 19.5$ corresponding to $ N \simeq 1710 $. These numbers compare
favourably with the spherically symmetric results. In that
case the limit found using the steepest descent method is $Q = 13.7$ corresponding to about $N \simeq 1200$, again lower than
the gaussian bound $Q = 16.7$ and then $N = 1460$. So the cigar-like trap allows 
to increase the maximum number of particles by  25\%. This is a small but significant increase. 
We have cheked by numerical solution of the gaussian equations of Ref. \cite{Victor2}
and numerical simulations of (\ref{NLS3}) that the cigar trap is the optimal one; i.e. there 
is no another parabolic trap 
configuration with better collapse-avoiding properties.

As stated in the introduction one would like to increase the number of particles in the  condensate as much as possible. This would allow
the control of a large coherent pulse of atoms. To do so
one could try to use a higher order soliton \cite{Miles}. 
However, for those 
solutions the shape performs complicated (but periodic) 
oscillations and develops high spatial and temporal gradients 
that probably would rule out the approximation and induce collapse if the 
order of the soliton is large enough.  It is also possible
to generate a soliton train where the solitons have different 
global phases so that the interaction 
between them is repulsive. This idea should work
for some situations allowing many particles to be
put in the ground state and will be elaborated in future work. 
Other possibility is to use a non--Gaussian fundamental mode for the 
transverse solution as in \cite{Dalfovo,Shi}. However the question of the stability of those  solutions under
general three-dimensional perturbations is not trivial and is the subject of current research. 
Finally, other possibilities proposed in the literature could be of use 
here such as  using two condensates
\cite{Thomas} or the control of 
the value of the scattering length \cite{Tiesinga}. 

\section{Conclusion}

We have found compact solutions of (\ref{NLS3}) that exist  
due to nonlinear effects even when the $z$ trapping potential is absent.
Joining this result with (\ref{Ehr}) we conclude that it is possible 
to control the motion of the condensate, which could propagate 
without distortion by using smoothly varying external potentials.
Thus, the atom cloud 
could be manipulated very easily, e.g., with
an atom guide. It is interesting and 
curious that this cigar-like packet could be transported in that rigid way
behaving like a quasiparticle. This behavior is specific of 
negative scattering length condensates and an advantage over the 
positive scattering length ones, which tend to fill all 
the available space due to the repulsive atom-atom interaction. 
Additionally we have pointed out that 
relaxing the trapping potential in one direction in current traps 
would allow to increase the number of particles that can 
be put into a negative scattering length condensate. 

We hope that this study will stimulate the experimental efforts
in performing BEC with negative scattering length and think that
 the soliton
solutions here studied will be of practical applicability 
in Bose-Einstein 
condensate ``engineering".  

\acknowledgements

V. M. P-G. and H. M.  acknowledge 
the hospitality of the
Institut f\"ur Theoretische Physik, 
Universit\"at Innsbruck, were part
of this work was done. We thank F. Dalfovo
for his help with the numerical simulations.
This work has been supported in part by the Spanish 
Ministry of Education and Culture under 
grants PB95-0389 and PB96-0534.

\appendix
\section*{Derivation of the 1-d NLSE by multiple scale analysis}

Here we will give the details of a more formal derivation of Eqs. 
(\ref{prima}) and (\ref{NLS2}) from (\ref{NLS3}) using multiple scale analysis
\cite{Bender}. Let us first choose the 

\begin{mathletters}
\label{scaling}
\begin{eqnarray}
u & = & \varepsilon^{1/2} u_0(\tau,\tau',x,y,z') +
 \varepsilon^3 u_1 + \varepsilon^{11/2} u_2 + ... \\
\lambda_z & = & \varepsilon^4 \lambda_z \\
z' & = & \varepsilon z \\ 
\tau' & = & \varepsilon^2 \tau \\ 
Q & = & \varepsilon Q_0 
\end{eqnarray}
\end{mathletters}

This scaling satisfies the desiderable property that the $L^2$ norm of $u$ is conserved and that the potential in $z$ is weaker than the nonlinear interaction (and the later weaker than the trasverse potential) when 
$\varepsilon \rightarrow 0$. Inserting (\ref{scaling}) into Eq. (\ref{NLS3})
we find 

\begin{eqnarray}
i \left(\frac{\partial}{\partial \tau} + \varepsilon^2 
 \frac{\partial}{\partial \tau'} \right) \left(\varepsilon^{1/2} u_0 + 
\varepsilon^3 u_1 + \varepsilon^{11/2} u_2  + ... \right)  =  \nonumber \\
 = \left[ -\frac{1}{2} \nabla^2_{xy} - \frac{1}{2} \varepsilon^2 \frac{\partial^2}{\partial z'^2} + \frac{1}{2} \left( \rho^2 + 
\varepsilon^4 \lambda_0^2
 \frac{1}{\varepsilon^2} z'^2 \right)  \right. \nonumber \\
 \left. 
- \frac{\varepsilon Q_0}{2} \varepsilon \left|u_0 + \varepsilon^{5/2} u_1 + ...
\right|^2 \right] 
\left(\varepsilon^{1/2} u_0 + \varepsilon^{3} u_1 + ... \right) \label{mult1}
\end{eqnarray}

We now separe Eq. (\ref{mult1}) in the different orders in $\varepsilon$

\begin{mathletters}
\begin{eqnarray}
O\left(\varepsilon^{1/2}\right): & i \frac{\partial u_0}{\partial \tau} = &
\left(-\frac{1}{2} \nabla_{xy}^2 + \frac{1}{2} \rho^2 \right) u_0 \label{oep12} \\
O\left(\varepsilon^{5/2}\right): & i \frac{\partial u_0}{\partial \tau'} = &
\left( -\frac{1}{2} \frac{\partial^2}{\partial z'^2} + \lambda_0^2 z'^2 \right) u_0 \nonumber \\ & & - \frac{Q_0}{2} |u_0|^2 u_0 \label{oep52} \\
O\left(\varepsilon^3\right): &  i \frac{\partial u_1}{\partial \tau} = &
\left( -\frac{1}{2} \nabla_{xy}^2 + \frac{1}{2} \rho^2 \right) u_1 \label{oep3} \\
O\left(\varepsilon^{5}\right): &  i \frac{\partial u_1}{\partial \tau'} = &
\left(-\frac{1}{2} \frac{\partial}{\partial z'^2} + \lambda_0^2 z'^2 \right) 
u_1 \nonumber \\ & & -\frac{Q_0}{2} 2 |u_0|^2 u_1 \label{oep5} \\
O\left(\varepsilon^{11/2}\right): &  i \frac{\partial u_2}{\partial \tau} = &
\left( -\frac{1}{2} \nabla_{xy}^2 + \frac{1}{2} \rho^2 \right) u_2 \\
O\left(\varepsilon^{15/2}\right): &  i \frac{\partial u_2}{\partial \tau'} = &
\left(-\frac{1}{2} \frac{\partial}{\partial z'^2} + \lambda_0^2 z'^2 \right) 
u_2 \nonumber \\ & & - \frac{Q_0}{2} 2 |u_0|^2 u_2 - \frac{Q_0}{2} u_1^2 u_0^* \label{oep9}
\end{eqnarray}
\end{mathletters}
where $\nabla_{xy}^2 = \frac{\partial^2}{\partial x^2} + \frac{\partial^2}{\partial y^2}$. 
Eq. (\ref{oep12}) implies that the trasverse profile of $u_0$ is given by the isotropic two dimensional 
harmonic oscillator equation and then $u_0$ can be choosen as $u_0 = \phi(x,y) \xi(z',\tau') e^{\nu \tau}$. Substituting into Eq. (\ref{oep12}), multiplying by $\phi^*$ and integrating over the trasverse coordinates $x,y$
we obtain 
\begin{equation}
i \frac{\partial \xi}{\partial \tau'} = \left( -\frac{1}{2} \frac{\partial^2}{\partial z'^2} + \lambda_0^2 z'^2 \right) \xi - \frac{Q_0}{4} |\xi|^2 \xi 
\end{equation}
This means that the longitudinal profile obeys the nonlinear Schr\"odinger 
equation. In the $\lambda_z=0$ case the solutions can be found analitically as discussed in Sec. II. 
Joining the longitudinal and the trasverse solution and changing back to the
nonscaled variables we find that the ground state solution 
has the form
\begin{equation}
\label{complete}
u(\rho,s,\tau) = \sqrt{\frac{Q}{4\pi}} \text{sech}\left(\frac{Qs}{8\pi}\right) e^{-\rho^2/2} e^{-i\nu_p \tau}
\end{equation}
at least to the first order in $\varepsilon \propto Q$. The corrections are given by Eqs. (\ref{oep3}-\ref{oep9}). It is easy to see that the equations have solutions $u_1 = u_2 = 0$, so the solution is determined at least to order $\epsilon^{11/2}$ by $u_0$. This is the reason why the ground state solution is close to the approximate profile given by Eq. (\ref{complete})
even in the nonperturbative region as discussed in Sec. II.

\begin{figure}
\caption{Sections of the ground state solution of Eq. (\ref{NLS3})
with $Q=5$ and different values of $\lambda_z$. From the innermost to the 
outermost curves the $\lambda_z$ parameter is $\lambda_z=0.4,0.2,0.0$
The dash-dot line corresponds to the theoretical prediction for $\lambda_z=0$
given by Eq. (\ref{soliton1}) while the dashed line corresponds to the 
Gaussian solution given by Eq. (\ref{gaussian}) for $\lambda_z=0.4$.
(a) $s$-section for $\rho=0$. (b) $\rho$ section for $s=0$.}
\label{Fig1}
\end{figure}

\begin{figure}
\caption{Sections of the ground state solution of Eq. (\ref{NLS3})
with $Q=10$ and $\lambda_z=0$. The dash-dot
line corresponds to the theoretical prediction for $\lambda_z=0$
given by Eq. (\ref{soliton1}) (a) $s$-section for
 $\rho=0$. (b) $\rho$ section for $s=0$.}
\label{Fig2}
\end{figure}

\begin{thebibliography}{99}
\bibitem{import} The fact that the external force is smoothly varying is important to avoid pulsations or even breaking of the condensate. Analysis of the soliton behavior under different external potentials can be seen in \onlinecite{Vazquez,potenti}.
\end{thebibliography}
\end{document}